# Adaptive Interface for Accommodating Colour-Blind Users by Using Ishihara Test

Muhammad Taha Khan

K142016@nu.edu.pk

Abu Zohran Qaiser

K142058@nu.edu.pk

Department of Computer Science

FAST-NUCES, Karachi.

## ABSTRACT

Imperative visual data frequently vanishes when color applications are seen by partially color blind users. A new method for adaptive interface for accommodating color blind users is presented. The method presented here has two sections: 1) test client perceivability by utilizing Ishihara plates. 2) change the interface color scheme to accommodate color blind users if necessary. We demonstrate how the method works via a simple interface and evaluate the efficiency of our method by experimenting it on 100 users.

## Author Keywords

Color blindness, Color Vision Deficiency (CVD), Adaptive interface.

## INTRODUCTION

Color blindness, or color vision deficiency (CVD), is known to be a noteworthy obstruction to compelling PC utilize [1]. Colour blindness or colour vision deficiency (CVD) affects around 1 in 12 men and 1 in 200 women worldwide. This means that for every 100 users that visits website or app, up to 8 people could actually experience the content much differently that you'd expect. Most colour-blind people are able to see things just as clearly as the rest of the population, the difference is their inability to distinguish red, green, or blue light. There are distinctive sorts of CVD and the level of CVD can fluctuate from individual to individual. The three main types of abnormal color vision system are called anomalous trichromatism, dichromatism and monochromatism. Anomalous trichromatism comes about when one of the key cones has had its peak affectability moved. The sorts are named protanomaly and deuteranomaly relying upon whether the L or M cones have been influenced. Anomalous trichromats impression of shading ranges from practically typical to dichromatic contingent upon the degree to which the flawed cone has had its peak affectability moved. Dichromatism is a serious type of CVD that outcomes when one of the principal cones is absent. Dichromats are delegated protanopes, deuteranopes or tritanopes, contingent upon whether the L, M or S cones are absent. Monochromatism is the severest form of CVD and is characterized by a total inability to distinguish colors [1,3,7]. Different methods are used for diagnosing color vision deficiency. Ishihara color test is regularly used to screen for congenital and procured red green inadequacies [2,6], and the attributes of the reactions may change with the severity of the imperfection [8].

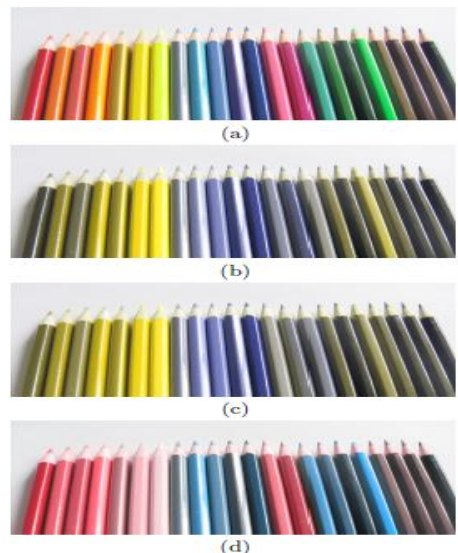

Figure 1: (a) a digital photograph of a set of 24 coloring pencils and computerized simulations depicting how it appears to (b) protanopes; (c) deuteranopes; and (d) tritanopes.

## RELATED WORK

A number of automatic adaptation algorithms have been proposed that modify content for CVD viewers. These methods formulate the problem of recolouring images for CVD viewers as one of optimization. The goal is to modulate colors in the image so that when they are viewed by a CVD person, the perceived difference between any pair of colors is of the same magnitude as that perceived by a normal color viewer. The optimization is complex, requiring the use of a small number of representative colors [1]. Our technique has impressive preferred standpoint as it is straightforward and reasonable. The speed of the technique and the mapping relies upon the Ishihara plates yield which is significantly speedier when contrasted with past strategies.

## DESIGN

To verify our proposed method, we designed the simple smartphone application named as 'LifeLine'. The interface of our application is shown in Figure 2. Lifeline is an application through which blood donors and acceptors can directly communicate with each other. At the starting of the application the users have to pass through the Ishihara test through which we choose that which color scheme is the most appropriate for the user. After that the application starts with most suitable color scheme for the particular user. User can Sign up as donor or acceptor. Donor can see ads posted by acceptors and acceptors can post an ad at whatever point essential.

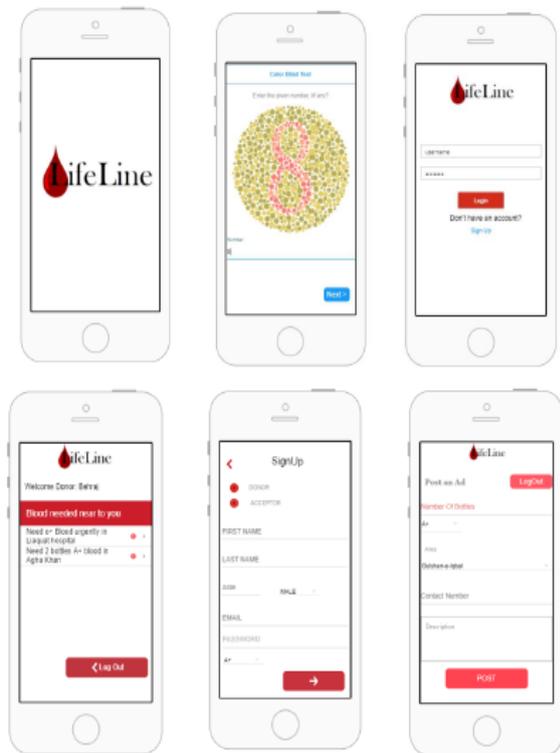

Figure 2: The interface of our application.

## METHOD

We led a preparatory assessment with 100 undergraduate students. All subjects with an age of 18 to 24 years participated in our experiment. All subjects with an age of 18 to 24 years took an interest in our analysis. Each subject began the Lifeline application and finished through the Ishihara color test. After that we approached each subject for his/her comments about the application shading scheme and interface outline and recorded that.

## RESULT

In the conducted experiment, it was found that only four out of hundred were colour blind. In that four three were suffering from deuteranopes and only one was suffering from protanopia. All the subjects suffering from colour-blindness found application colour scheme satisfactory and visible.

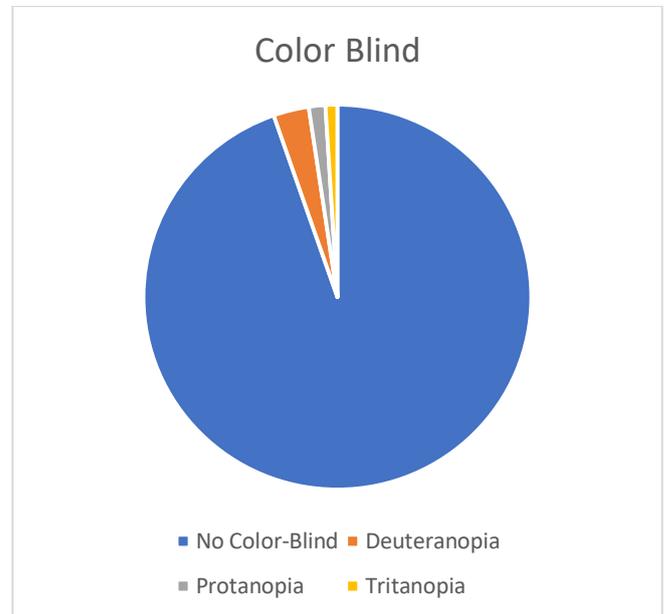

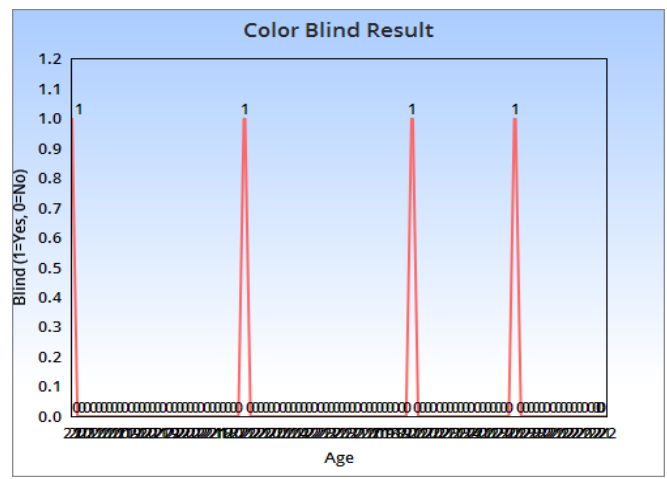

## CONCLUSION

The paper presented a new and simplified way to design an interface that will also support colour blind users. We evaluated our research via a simple interface and evaluated our method on the basis of subject's comments. The method is therefore successful, at least in terms for those who doesn't know that they are colour blind.